\documentclass[final,5p,times,authoryear]{elsarticle}

\journal{Chemical Physics Letters}

\begin{document}
\sloppy
\begin{frontmatter}
\title{From planes to bowls: photodissociation of the bisanthenequinone cation}

\author{Tao Chen$^{1,2}$, Junfeng Zhen$^{2,3}$, Ying Wang$^{1}$, Harold Linnartz$^{3}$, Alexander G. G. M. Tielens$^{2}$} 
\fntext[]{Corresponding author: Tao Chen (chen@strw.leidenuniv.nl)}

\address{$^1$Department of Theoretical Chemistry and Biology, School of Biotechnology, Royal Institute of Technology, 10691, Stockholm, Sweden\fnref{label1}}
\address{$^2$Leiden Observatory, Leiden University, PO Box 9513, NL 2300 RA Leiden, the Netherlands\fnref{label2}}
\address{$^3$Sackler Laboratory for Astrophysics, Leiden Observatory, Leiden University, P.O.\ Box 9513, NL 2300 RA Leiden, The Netherlands\fnref{label3}}

\begin{abstract}
We present a combined experimental and theoretical study of the photodissociation of the bisanthenequinone (C$_{28}$H$_{12}$O$_2$) cation, Bq$^+$. The experiments show that, upon photolysis, the Bq$^+$ cation does not dehydrogenate, but instead fragments through the sequential loss of the two neutral carbonyl groups, causing the formation of five-membered carbon cycles. Quantum chemical calculations confirm this Bq$^+$ $\rightarrow$ [Bq - CO]$^+$ $\rightarrow$ [Bq - 2CO]$^+$ sequence as the energetically most favorable reaction pathway. For the first CO loss, a transition state with a barrier of $\sim$ 3.2 eV is found, substantially lower than the lowest calculated H loss dissociation pathway ($\sim$ 4.9 eV). A similar situation applies for the second CO loss channel ($\sim$3.8 eV vs. $\sim$4.7 eV), but where the first dissociation step does not strongly alter the planar PAH geometry, the second step transforms the molecule into a bowl-shaped one.
\end{abstract}


\end{frontmatter}

\section{Introduction}
Bisanthenequinone (Bq) belongs to the family of the polycyclic aromatic hydrocarbon (PAH) quinones. These molecules are known products in the photo-oxidation of environmentally relevant PAHs \citep{alam, garcia}. During incomplete combustion processes quinones are released into the atmosphere \citep{iinuma, layshock, valavanidis}. They are also used in (in)organic synthesis, as oxidizing agent \citep{march} or because of their pharmacological relevance \citep{patai,liu}. They may also be of astronomical relevance \citep{tielens2008}, the motivation for the present study.

The vibrational signatures of PAHs dominate the mid-infrared spectra of many objects in space and are key contributors to the energy and ionization balance of the gas \citep{tielens2008}. Interstellar PAHs are assumed to form in processes akin to soot formation in the cooling ejecta of carbon-rich Red Giant stars as they flow from the stellar photosphere into the interstellar medium \citep{frenklach,cherchneff1992}. Subsequently, they are further processed for millions of years by photons of the interstellar radiation field \citep{tielens2013}. Driven by this astrophysical interest, experiments focusing on the photoexcitation of PAHs have attracted much interest in recent years \citep{zhen2014, zhen2016} (and references therein). A number of different processes can take place, varying from sequential fragmentation \citep{ekern1998,west2012dissociation,zhen2014}, isomerization \citep{dyakov2006ab, johansson2011unimolecular,solano2015complete,simon2017dissociation,trinquier2017pah} and ongoing ionization \citep{holm2011,zhen2015laboratory}. Dedicated studies of the involved dissociation channels provide information on the molecular dynamics at play and this is interesting, both from an astronomical and physical chemical point of view \citep{holm2011,chen2015}. Particularly processes changing the nature of the carbon skeleton have been the topic of recent studies. In photodissociation regions (PDRs) in space, (large) PAHs (with more than 50 C-atoms) are considered starting points in the formation of other species, including fullerenes, carbon cages and smaller hydrocarbon chains \citep{pety2005,berne2012,zhen14,west2012dissociation}. Also PAHs with functional side groups may be important. Inside molecular clouds, PAHs are expected to be trapped in low temperatures ($\sim$10 K) ice mantles \citep{guennoun2011jpca,guennoun2011pccp}, consisting mainly of H$_2$O with traces of CH$_3$OH, CO$_2$, CO, and NH$_3$. Photolysis of these complex ice mixtures is known to functionalize PAHs with alcohol (-OH), ketone ($>$C=O), amino (-NH$_2$), methyl (-CH$_3$), methoxy (-OCH$_3$), cyano/isocyano (-CN, -NC), and carboxyl (-COOH) groups \citep{bernstein2002,cook2015}. When molecular cloud ices are exposed to the strong radiation field of a newly formed massive star in a PDR, ice molecules can be returned to the gas phase through various processes \citep{tielens2013}. Photolysis of these newly formed PAHs with functionalized side-group additions, like methyl, methoxy, hydroxyl or carbonyl groups may play a comparable important role \citep{bernstein2002}.

Quinones mainly dissociate through consecutive losses of carbonyl units \citep{beynon59, proctor, pan2008}. Their fragmentation mechanisms are well established for small species, e.g. substituted 1,4-naphthoquinone \citep{becher, mari, stensen, pan2008} and substituted anthraquinone \citep{proctor, beynon60}. Experimental and theoretical studies of larger quinones have not been described in previous reports. In this work, we study the photodissociation of the large quinone cation (4,11-bisanthenequinone, CAS No. 475-64-9). Both experimental and theoretical studies are carried out for understanding the effect of acetone site-substitution on photostability and photoreactivity. The experiments are conducted using quadrupole ion trap time-of-flight (QIT-TOF) mass spectrometry. Quantum chemical calculations are performed to explore the dissociation pathways and resulting geometry changes.

The article is organised as follows: Section 2 and 3 describe the experimental setup and computational methods used for this study. Section 4 shows the experimental and theoretical results, and discusses the astrophysical relevance. The conclusions follow at the end. 

\section{Experimental setup}
The photofragmentation experiments on Bq cations are conducted with i-PoP, our instrument for photodissociation of PAHs \citep{zhen14} that comprises a commercial quadrupole ion trap time-of-flight system. A typical experiment works in the following way: commercially available Bq powder (purity higher than 99.0 \% from Kentax) is heated to $\sim$500 K in a small oven. Subsequently, the evaporated molecules are ionized by an electron gun and the ions are guided into the ion trap. Once the ions are trapped, a stored waveform inverse Fourier transform (SWIFT) excitation technique \citep{doroshenko} is used to isolate a specific range of mass/charge (m/z) species. After a short time delay (typically $\sim$0.2 s), the ion cloud is cooled down to room temperature ($\sim$298 K) through collisions with He buffer gas that is continuously added to the trap. Subsequently, the ion cloud is irradiated with light pulses generated by a tunable dye laser (LIOP-TEC, Quasar2-VN) pumped by a Nd:YAG laser (DCR-3, Spectra-Physics), operated at 10 Hz. A solution of DCM dye is used to produce 626-nm-light as well as 312 and 208 nm radiation through doubling and tripling of the original light. The fragments and intact molecules after irradiation with several photons are accelerated by a negative square pulse to transfer them from the trap to the field-free TOF region, where the corresponding mass signals are detected using a microchannel plate (MCP) detector. The full system operates at pressures of the order of 10$^{-8}$ mbar or better. 

\begin{figure}[t]
  \centering
  \includegraphics[width=1.0\columnwidth]{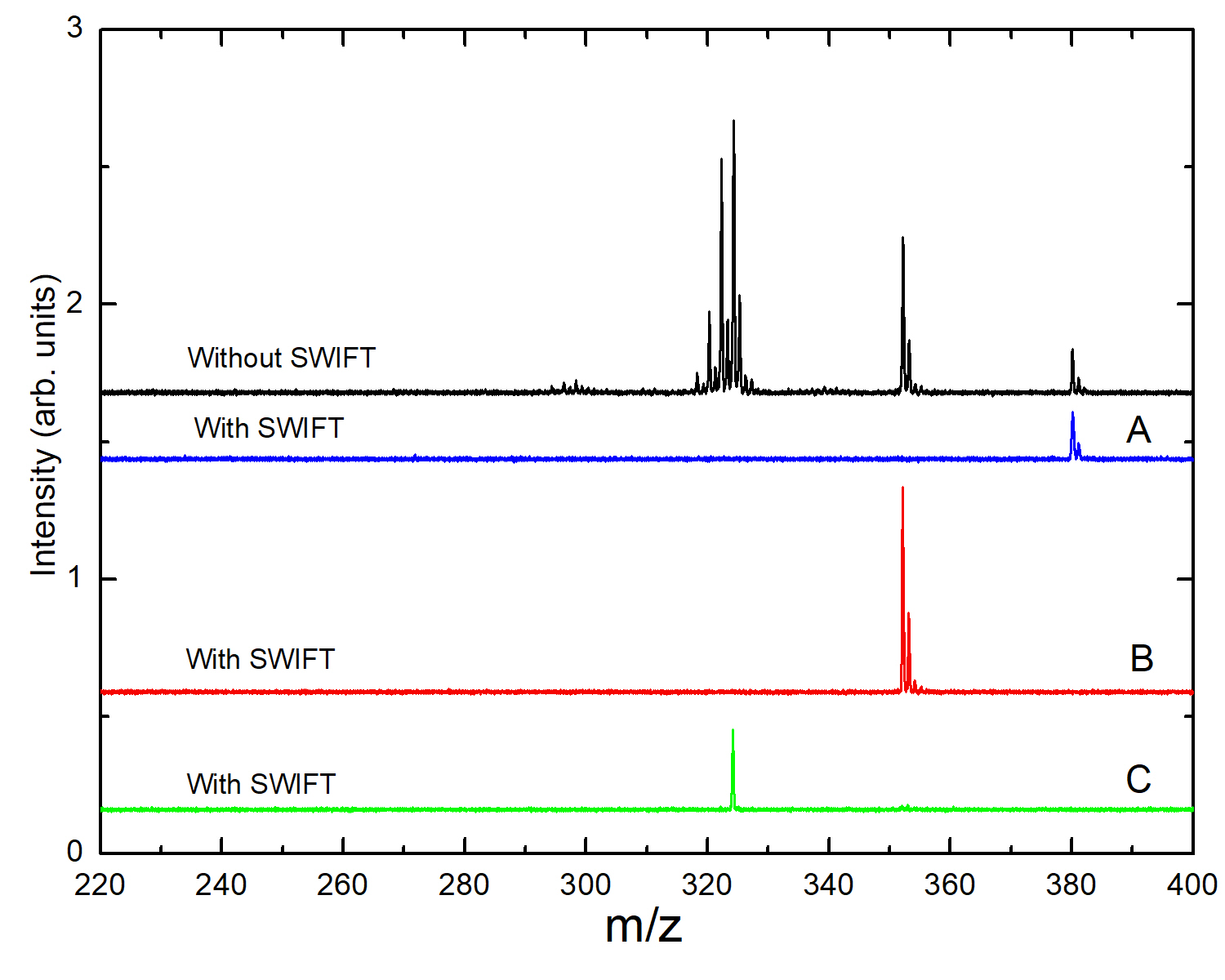}
  \caption{The electron impact induced mass spectrum of Bq cation, before (black curve) and after (blue (A), red (B) and green (C) curve) SWIFT signal isolation. The blue curve at mass $\sim$380 amu corresponds to the Bq$^+$. The red curve at mass $\sim$352 amu corresponds to one carbonyl loss from Bq$^+$ ([Bq - CO]$^+$). The green curve at mass $\sim$324 amu corresponds to the two carbonyl losses from Bq$^+$ ([Bq - 2CO]$^+$). The small peaks on the right side of the main peak in graphs A and B are isotopic contributions from $^{13}$C or $^{18}$O, but there is not enough oxygen incorporated to generate detectable amounts of isotopomers given the apparent signal-to-noise ratio: only $^{13}$C enriched species are actually visible in the mass spectra. SWIFT isolation effectively reduces isotopes in the mass spectrum of graph C, i.e. [Bq - 2CO]$^+$.}
\end{figure}

\begin{figure*}[t]
  \centering
  \includegraphics[width=1.0\textwidth]{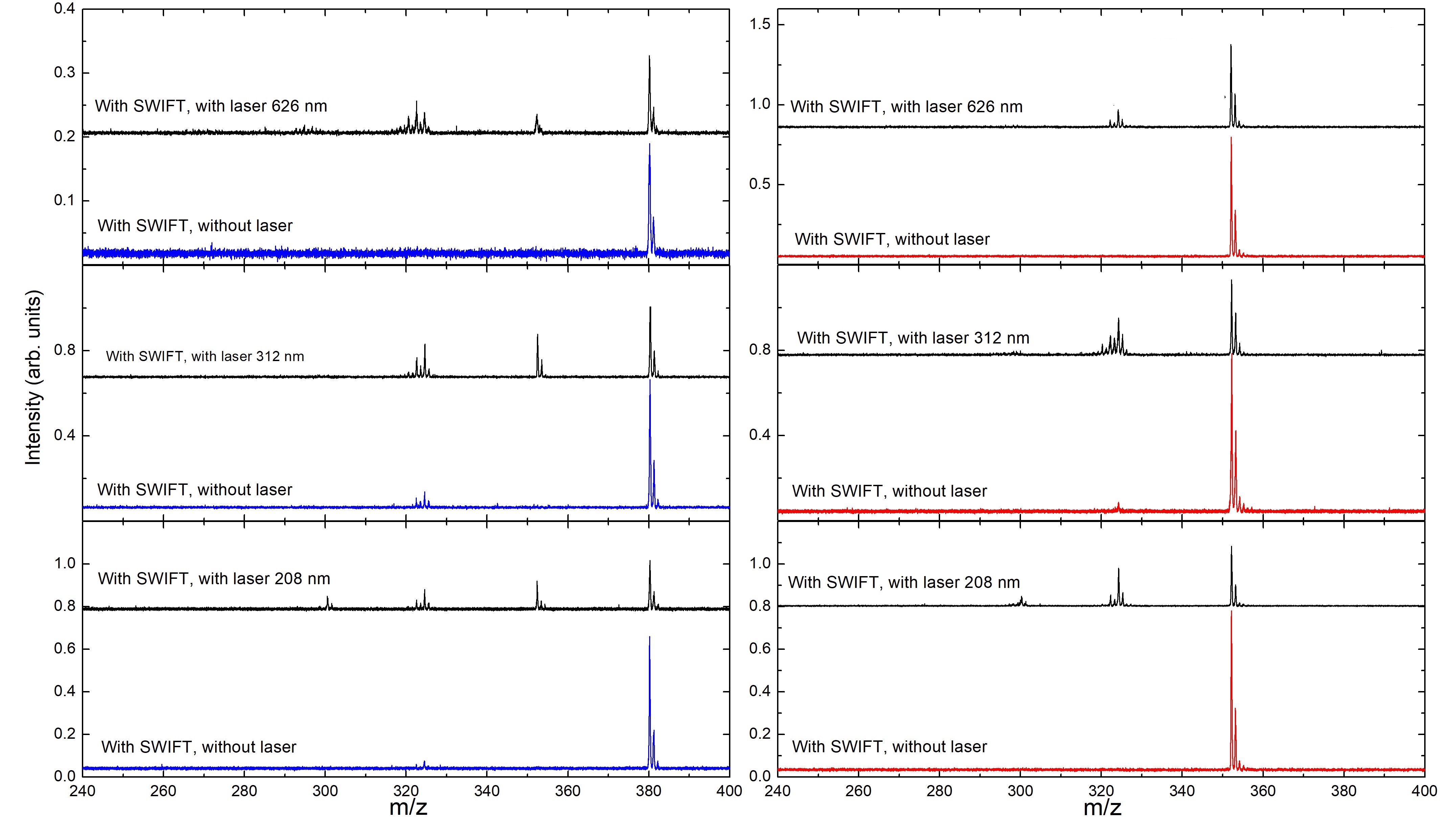}
  \caption{Mass spectrum of Bq$^+$ (left panels) and [Bq - CO]$^+$ (right panels) before (lower graph in each panel) and after irradiation (upper graph in each panel) by 626 nm (top panels), 312 nm (middle panels) and 208 nm (bottom panels). The blue (red) curves represent SWIFT selected Bq$^+$ ([Bq - CO]$^+$) without laser irradiation. The black curves show the effect upon irradiation. No proof for dehydrogenation is found, meaning that CO is the dominant dissociation channel for Bq$^+$ and [Bq - CO]$^+$ (see text).}
\end{figure*}

\begin{figure}[t]
  \centering
  \includegraphics[width=1.0\columnwidth]{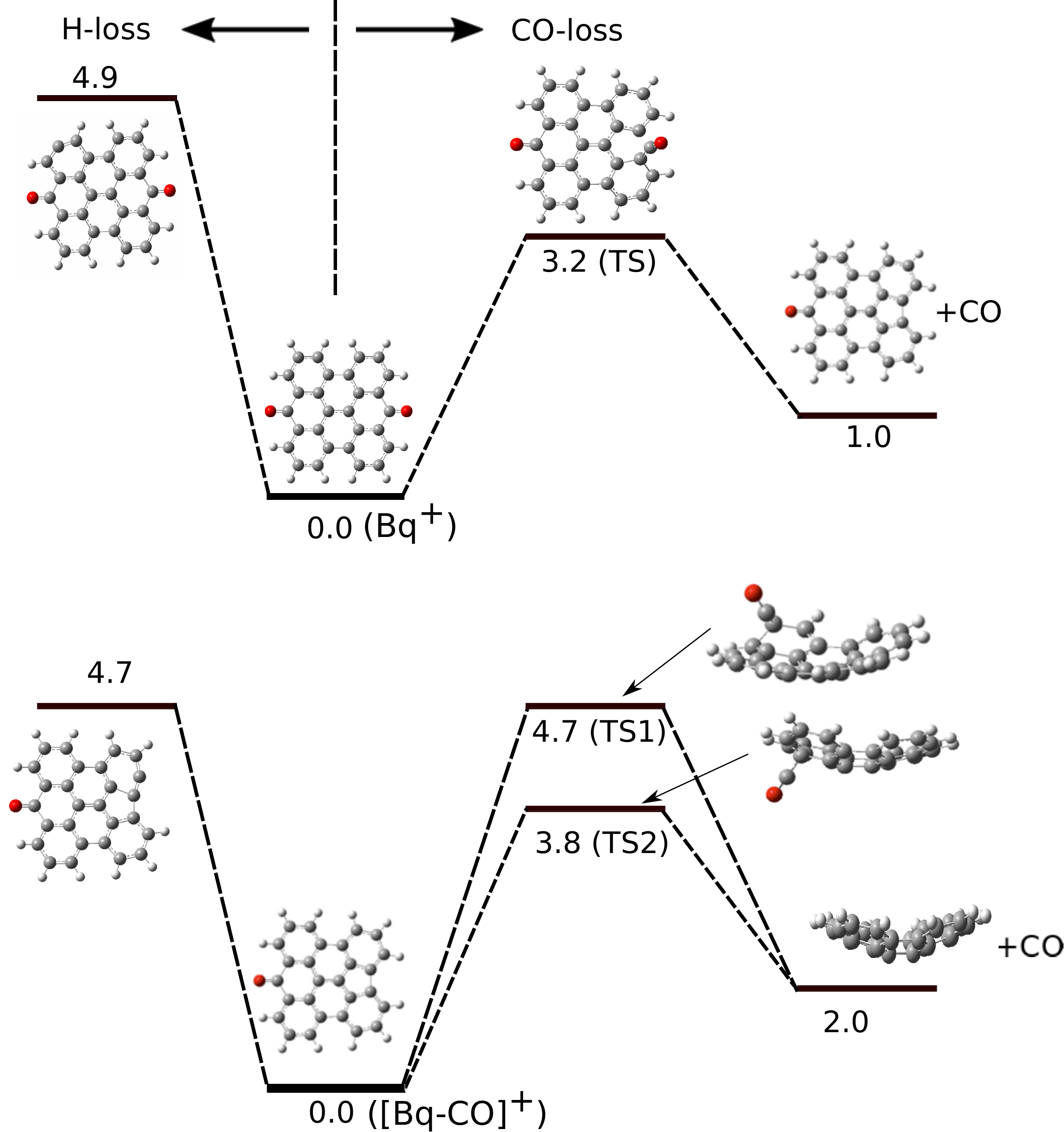} 
  \caption{Calculated dissociation energies and reaction barriers for H losses (three different H losses in Bq$^+$ and six in [Bq - CO]$^+$, here only the energetically most favorable channel is shown) and CO losses from Bq and [Bq - CO]$^+$ cation. The barrier for CO loss from Bq$^+$ is significantly lower than the lowest dissociation energy of H-loss. Two reasonable barriers are found for the second CO loss from [Bq - CO]$^+$, one is comparable with H-loss, the other one is 0.9 eV. All values are given in eV.}
\end{figure}

\section{Computational methods}
Our theoretical calculations are carried out using density functional theory (DFT). The dissociation energies, transition state energies and dipole moments presented in this work are calculated using the hybrid density functional B3LYP \citep{becke,lee} as implemented in the Gaussian 16 program \citep{frisch2016gaussian}. All structures are optimized using the 6-311++G(2d,p) basis set. The vibrational frequencies are calculated for the optimized geometries to verify that these correspond to minima or first-order saddle points (transition states) on the potential energy surface (PES). We have taken the zero point vibrational energy (ZPVE) into account. The ZPVE values are scaled by the empirical factor 0.965 to correct for anharmonic effects \citep{andersson2005new}. It should be noted that these corrections will not influence our conclusions regarding the dissociation energies and transition states calculations. The calculated energies for the small [6-31G(d)] and the larger basis set [6-311++G(2d,p)] are very similar, and thus, only the results from the larger basis set calculations are presented here. The PES of all possible dissociation pathways is scanned for transition state calculations. Intrinsic reaction coordinate (IRC) calculations \citep{fukui,dykstra} are performed to confirm that the transition state structures are connected to their corresponding local PES minima. In all cases we have only considered the ground state PES, as the non-radiative decay for such a large molecule is very rapid: $\sim$ 10$^{-12}$ s \citep{vierheilig1999femtosecond, zewail2000femtochemistry}. Only the lowest spin state is considered, which seems to be a well-justified approach to predict reasonable energies in comparison to the experimental results \citep{holm2011}.

\section{Results and discussion}
Figure 1 shows the mass spectrum of the trapped Bq$^+$ after electron impact ionization of the Bq precursor species. The top mass spectrum is obtained without laser irradiation and without SWIFT signal isolation. It illustrates that the parent cations experience substantial fragmentation due to the electron impact ionization. The improvement in mass purity becomes clear from the lower three curves, where mass peaks at m/z $\sim$380 for the precursor species (Bq$^+$), at $\sim$352 for the [Bq - CO]$^+$ and at $\sim$324 for the [Bq - 2CO]$^+$ are shown using the SWIFT pulse isolation. The mass spectra of Bq$^+$ and [Bq - CO]$^+$ show that the relatively small isotopic contributions from $^{13}$C or $^{18}$O, at m/z $\sim$381 \& 382 and 353 \& 354 are further suppressed by the SWIFT pulse. Isolation of the parent mass peak, excludes $^{13}$C contributions in the mass spectrum of the [Bq - 2CO]$^+$.

Figure 2 shows the resulting mass-spectrum for Bq$^+$ irradiation (left panels) and [Bq - CO]$^+$ irradiation (right panels) for three different wavelengths: 626 nm (upper panels), 312 nm (middle panels), and 208 nm (bottom panels). In all panels the SWIFT reference mass spectrum (without laser) is included as well, in blue for Bq and in red for [Bq - CO]$^+$. Clearly, due to multi-photon absorption in experiments for the three different wavelengths, only minor changes have been observed in the resulting mass spectra. For this reason, in the remaining of this article, we only will present and discuss the 626 nm data. The fragmentation through the loss of two CO-units is clearly the dominant dissociation pathway. Once this process has ended, the [Bq-2CO]$^+$ starts behaving like a regular PAH \citep{chen2015}. In Figure 2 (left bottom panel) it can be seen that a C$_2$-loss channel results in m/z signal at 300, and corresponding C$_2$H$_2$-loss in m/z signal at 298.  

As shown in Figure 1, the precursor molecule is located at m/z $\sim$380. Two main photofragments are found with peaks at m/z $\sim$352 and 324, very similar to the electron impact induced fragmentation. The separation between precursor and first fragment peak as well as between first and second fragment peak amounts to 28 Da mass differences. This corresponds to one oxygen plus one carbon (m/z = 28), fully consistent with the sequential loss of two CO units. Moreover, no other peaks are observed at Bq$^+$ and [Bq - CO]$^+$ except those due to isotopes. Hydrogen loss channels are not found. Our interpretation of these observations is that Bq photofragmentation is governed by loss of the two CO units only, transforming Bq into [Bq - CO]$^+$ and [Bq - 2CO]$^+$. It is interesting to note that this CO-loss channel is clearly preferred above dehydrogenation, i.e., fragmentation through H-atom loss. For many regular large PAHs, i.e., PAHs without side groups, H loss and 2H/H$_2$ loss are dominant dissociation channels \citep{chen2015}. This observation is remarkable although not fully unexpected as the same observation was reported for smaller species like 1,2-naphthoquinone and others \citep{pan2008}. Also in the case of PAHs with other side group additions, like methoxy and methyl groups, dissociation pathways other than H-losses were reported \citep{zhen2016}. 

In order to further understand the details of the dissociation process, quantum chemical calculations are performed. Figure 3 shows the calculated dissociation energies and reaction barriers for H loss and CO loss from Bq$^+$ and [Bq - CO]$^+$. Given its molecular geometry, Bq$^+$ has three possible ways to lose the first hydrogen, These have been considered and our calculations show that the lowest dissociation energy for H loss from Bq$^+$ is about 4.9 eV (see Figure 3 for the position of such H atom in Bq$^+$), which is very similar to values found for regular PAHs \citep{chen2015}. The barrier for one CO loss from Bq$^+$ is calculated to be $\sim$3.2 eV which is significantly lower than the energy needed to release an H-atom. The imaginary frequency for the first and highest transition state ($\sim$3.2 eV) is about -550 cm$^{-1}$. Following the first barrier several shallow intermediate states are found before CO completely dissociates from the [Bq - CO]$^+$ plane (final state). However these intermediate states do not affect the general trend for the dissociation process found here and therefore only the first transition state and final state are shown in Figure 3. This explains why no H loss is seen around peak m/z $\sim$382. Clearly, the loss of one CO-unit is highly favored above that of an H-atom, which is fully consistent with the experimental result. 

The experiments indicate that a similar situation applies for the photodissociation of [Bq - CO]$^+$. In Figure 3 it is shown that [Bq - CO]$^+$ has six different dissociation channels for H losses. The lowest dissociation energy for H loss from [Bq - CO]$^+$ is about 4.7 eV (see Figure 3 for the position of such H atom in Bq$^+$). Two possible dissociation pathways are found for the second CO loss from [Bq - CO]$^+$, with barriers of $\sim$3.8 eV and $\sim$4.7 eV, and with imaginary frequencies about -435 cm$^{-1}$ and -301 cm$^{-1}$. During this process, the planar structure of the molecule distorts into a bowl structure and these two transition states correspond to a CO molecule above and below the bowl (see Figure 3 for a structural representation). The energy for TS1 (4.7 eV) is comparable to the value calculated for an H loss in [Bq - CO]$^+$. As no H loss is observed in the mass spectrum of [Bq - CO]$^+$, it is unlikely that this provides an active fragmentation channel. The other channel is lower in energy and prefers CO loss above dehydrogenation, explaining why no H loss around the peak at m/z $\sim$352 is found. Interestingly, whereas the [Bq - CO]$^+$ barely changes its molecular structure upon CO loss, we find that after removing the second CO group, Bq$^+$ loses its planar geometry and starts bending to form a bowl shaped PAH. This is discussed below.

\begin{figure*}[t]
  \centering
  \includegraphics[width=1.0\textwidth]{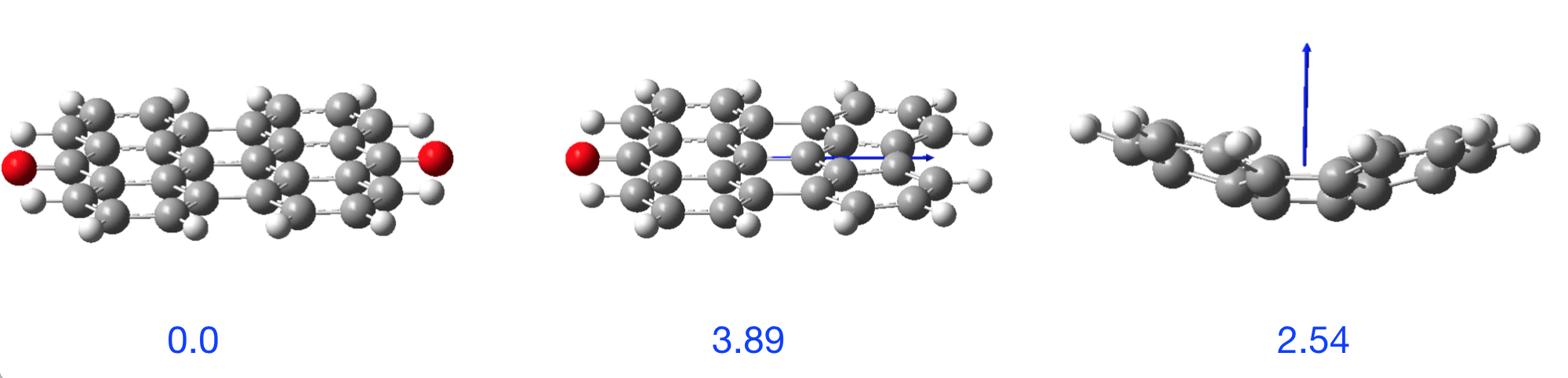} 
  \caption{Calculated dipole moments for optimized structures of Bq$^+$, [Bq - CO]$^+$, and [Bq - 2CO]$^+$ cations. Due to the change of structure, permanent dipole moments are induced. Arrows on each structure indicate the direction of the total dipole moment, and the corresponding values are given beneath each molecule in units of Debye.}
\end{figure*}

Figure 4 shows the calculated structure changes of Bq$^+$, [Bq - CO]$^+$, and [Bq - 2CO]$^+$ and their corresponding dipole moments. The bending of the aromatic structure presently observed results from the conjunction of (i) the particular arrangement of aromatic cycles in the structure of Bq$^+$ and (ii) the specific position of each CO group, not only with respect to the structure but also relative to each other. For [Bq-CO]$^+$ the formation of a pentagon at the periphery of this species after the loss of the first CO group does not lead to a large scale restructuring. However, after formation of the second pentagon, the molecular structure has to turn into a bowl-like geometry. We note that for this species the two pentagons are only separated by one C-C bond which will make it easier to initiate bowl formation, compared to larger PAHs. The original Bq$^+$  is fully symmetric and has no dipole. The highly asymmetric intermediate ([Bq - CO]$^+$) has a dipole moment of 3.89 D, situated in the plane, and the final product ([Bq - 2CO]$^+$) ends up with a 2.54 D dipole moment pointing inwards. It is clear that dissociation of CO and the curling of a PAH plane induces a dipole moment which would make such a molecule, in principle, detectable by radio astronomy \citep{lovas2005interstellar}. 

The work described here shows a rather specific pathway towards pentagon formation in PAHs. When exposed to UV photons, experiments show that large PAHs can quickly loose all of their peripheral hydrogens and subsequently isomerize to cages and fullerenes, typically after losing C$_2$-units from the carbon skeleton. It was shown that C$_{60}$ can form from C$_{66}$H$_{26}$ \citep{zhen2014} in line with the suggestion that the increased abundance of C$_{60}$ in photodissociation regions can be ascribed to complete H-loss followed by C-loss creating pentagons and initiating fullerene formation \citep{chuvilin2010direct, berne2012}. The study reported here shows that for the studied Bq, pentagon formation and curling of the planar PAH structure can commence before H-stripping starts, triggered by the dissociation of both side groups. The size of the involved molecule and the position of the side groups, obviously, are relevant parameters in this process. Our result also has some similarities with recent work by de Haas et al \citep{haas} showing that the loss of a HCN-fragment in nitrogen containing PAHs offers a facile pathway towards pentagon formation. We also mention the formation of fulvene-type isomers leading to bowl-shaped structures as theoretically predicted by \citep{trinquier2017pah}. As a consequence, the relevance of pentagon formation before H-stripping - as discussed in other recent work - may be a more general process taking place in space, but this will remain to be sorted out in future studies.

\section{Conclusion}
Experiments and quantum chemical calculations are performed for understanding the photodissociation processes of Bq cations. The mass spectrum of Bq cation upon laser irradiation with three different wavelengths (626 nm, 312 nm, and 208 nm) presents clear evidence for pure neutral CO-losses, i.e., no other fragments are detected as a first or second dissociation product. The CO-losses are highly favorable for all tested wavelengths. Our quantum chemical calculations reveal that for Bq the transition barrier for CO-loss is only $\sim$ 3.2 eV, which is much lower than alternative routes also including dehydrogenation. The loss of the second CO-unit costs more energy (3.8 eV), as a geometry change is involved with the introduction of the second pentagon into the molecular structure. The geometry change causes the planar PAH structure to change into a bowl-shaped form. This process from planes to bowls is shown in Figure 4 and in the graphical abstract.

The present study describes a fairly accessible pathway to pentagon formation in PAHs. We expect that CO loss from other PAH quinones will also lead to pentagon formation and possibly, depending on molecular size, bowl-shaping, creating a permanent dipole moment that makes these molecules visible for radio astronomy. 

\section{Acknowledgments}
This work is supported by Swedish Research Council (Contract No. 2015-06501). Facility is supported by the Swedish National Infrastructure for Computing (SNIC). We acknowledge the European Union (EU) and Horizon 2020 funding awarded under the Marie Sk\l{}odowska Curie action to the EUROPAH consortium, grant number 722346. Studies of interstellar PAHs at Leiden Observatory are supported through a Spinoza award.

\end{document}